\documentclass[11pt]{article}
\usepackage{jheppub}
\usepackage{amsmath}
\usepackage{amssymb}
\usepackage{braket}
\usepackage{bbold}
\usepackage{pifont}

\newcommand{\beqn}{\begin{eqnarray}}
\newcommand{\eeqn}{\end{eqnarray}}
\newcommand{\dd}{\mathrm{d}}
\newcommand{\nn}{\nonumber}

\newcommand{\tr}{\mathsf{{\scriptscriptstyle T}}}
\newcommand{\p}{\partial}

\newcommand{\gmn}{g_{\mu\nu}}
\newcommand{\fmn}{f_{\mu\nu}}

\newcommand{\JJ}{\hat{\mathrm{I}}}
\newcommand{\mT}{\mathsf{{\scriptscriptstyle T}}}

\title{Absence of ghost in a new bimetric-matter coupling}
\author{S.F.~Hassan,}
\author{Mikica Kocic,}
\author{Angnis~Schmidt-May}
\affiliation{Department of Physics \& 
        The Oskar Klein Centre,\\
        Stockholm University, AlbaNova University Centre, 
        SE-106 91 Stockholm, Sweden}
\emailAdd{fawad@fysik.su.se}
\emailAdd{mikica@kth.se}
\emailAdd{angnis.schmidt-may@fysik.su.se}

\abstract{Interactions in bimetric theory, which can describe gravity
  in the presence of an extra spin-2 field, are severely constrained
  by the requirement of the absence of the Boulware-Deser
  ghost instability. Recently an interesting new matter coupling was
  proposed in terms of a composite metric but it was claimed to
  reintroduce the ghost. In this paper we carry out a nonlinear
  Hamiltonian analysis of this new matter coupling and show that it is 
  indeed ghost-free. The analysis involves using a new set of
  variables that naturally appear in the relation between the metric
  and vielbein formulations of bimetric theory. In terms of these
  variables we show that the new matter coupling does not reduce the
  number of constraints in bimetric theory and hence does not
  reintroduce the Boulware-Deser ghost.}  

\toccontinuoustrue
\begin{document} 
\maketitle
\flushbottom

\section{Introduction}

Attempts to generalize linear Fierz-Pauli theory~\cite{Fierz:1939ix}
and construct nonlinear interactions for massive spin-2 fields
remained unsuccessful for decades due to the notorious appearance of
the Boulware-Deser ghost instability~\cite{Boulware:1972zf,
  Boulware:1973my}. Only a few years ago the development of new
approximation methods in \cite{ArkaniHamed:2002sp, Creminelli:2005qk}
lead to the proposal of a candidate action for ghost-free nonlinear
massive gravity \cite{deRham:2010ik, deRham:2010kj}. The formulation
of this type of theory requires the presence of a second metric tensor
(a ``background metric'') which, in this first approach, was taken to
be of Minkowski form. In \cite{Hassan:2011hr} it was shown that this
model indeed avoids the Boulware-Deser ghost and therefore gives the
first consistent description of nonlinear spin-2 self-interactions. A
reformulation and generalization to arbitrary background metrics
appeared in \cite{Hassan:2011vm}, for which the absence of ghost has
also been demonstrated \cite{Hassan:2011tf, Hassan:2011ea,
  Hassan:2012qv}. Moreover, it turned out that the second metric can
have its own dynamics without spoiling the consistency of the
theory~\cite{Hassan:2011zd, Hassan:2011ea}. This ghost-free bimetric
theory, whose spectrum around Einstein solutions has been analyzed in
\cite{Hassan:2012wr}, describes nonlinear interactions of a massive
and a massless spin-2 field. It can be regarded as a theory of
gravitational interactions in the presence of an extra spin-2 field. 

The only known consistent matter couplings in bimetric theory so far
have been of the same form as in general relativity, but now there are
two metrics that could possibly interact with two different types of
matter. From the ghost proof in~\cite{Hassan:2011zd}, it is clear that
both of the metrics can couple independently to different matter
sources without introducing inconsistencies. Recent work has shown
that a problem occurs in general when both metrics couple to the same
matter~\cite{Yamashita:2014fga, deRham:2014naa}. In this case, the
constraint that removes the Boulware-Deser ghost in bimetric theory is
destroyed and the fatal instability reappears. These findings are in
agreement with earlier attempts to couple matter to a certain
combination of the two metrics, which possesses massless fluctuations
around maximally symmetric solutions~\cite{Hassan:2012wr}. 

Recently, the authors of \cite{deRham:2014naa} identified another 
particular combination of the metrics that leads to interesting
consequences when coupled to matter. As for the ghost issue, they
showed that the new couplings were ghost-free in some simplifying
limits, but from a more detailed Hamiltonian analysis in a
perturbative setup, they concluded that the ghost instability was
indeed present sourced by the new matter couplings.\footnote{Shortly 
after, an independent group proposed the same kind
  of coupling~\cite{Noller:2014sta}, based on a simple, necessary but
  not sufficient condition for the absence of ghost.} If true, the
proposed matter coupling would reintroduce fatal instabilities and
thus would be of little use for phenomenological applications.
Although \cite{deRham:2014naa} argued that the theory could still be
treated as a valid effective theory with a cut-off below the ghost
mass, it is not obvious that such an interpretation is valid. Since
ghost modes can create energy from the vacuum, they are produced in
interactions with healthy fields even in the absence of external
energy \cite{Woodard:2006nt,Sbisa:2014pzo}. Spontaneous vacuum decay
into heavy ghost modes can release energy above the cut-off scale,
rendering an effective field theory description invalid.

In this paper we carry out a detailed non-linear Hamiltonian analysis
of the matter couplings proposed in \cite{deRham:2014naa} and show
that they are free of the Boulware-Deser ghost. The analysis involves
a modification of the construction that was used to prove the absence
of ghost is massive gravity and bimetric theory 
\cite{Hassan:2011hr, Hassan:2011tf, Hassan:2011zd}, and
naturally appears in the context of the connection between the metric
\cite{Hassan:2011zd} and vielbein \cite{Hinterbichler:2012cn}
formulations of bimetric theory.  

This paper is organized as follows. In section \ref{revbim} we review
the bimetric action and its ghost analysis, explaining why a main
step in the the analysis is guaranteed to work. In section \ref{adm2}
we introduce a new set of bimetric variables that simplifies the
expressions and makes the analysis feasible. The absence of ghost in
the proposed matter coupling is demonstrated in section
\ref{consproof}. For completeness, in section \ref{symvielbein} we
show that the new variables introduced possess an interesting
interpretation in the vielbein formulation. Finally, our results are
discussed in section \ref{discuss} and some mathematical details are
relegated to the appendices.

\section{Review of bimetric theory}\label{revbim}
In this section we first review bimetric theory and its matter
couplings and then provide a simplified overview of the ghost analysis
in the Hamiltonian framework.  
\subsection{The ghost-free bimetric action and matter couplings}
 
The ghost-free action for the two spin-2 fields $\gmn$ and $\fmn$ is
given by \cite{Hassan:2011zd,Hassan:2011ea}, 
\beqn
S=\int\dd^4x\,\Big[m_g^2\sqrt{g}\,R(g)+m_f^2\sqrt{f}\,R(f)-2 m^4\sqrt{g}
\sum_{n=0}^4\beta_n e_n\big(\sqrt{g^{-1}f}\big)\Big]\,.
\label{action}
\eeqn
Here, $m_g$ and $m_f$ are the Planck masses for the two metrics, $m$
is a mass scale and $\beta_n$ are interaction parameters. $e_n(X)$
denote the elementary symmetric polynomials of the eigenvalues of the
matrix $X$. They appear in the expansion of the determinant,  
\beqn
\det (\mathbb{1}+ X)=\sum_{n=0}^4e_n(X)\,.
\label{det-e}
\eeqn
The interaction potential originated in the massive gravity context
\cite{deRham:2010kj}, as reformulated in \cite{Hassan:2011vm}. The
dependence on the square-root matrix $X=\sqrt{g^{-1}f}$, defined through
$\big(\sqrt{g^{-1}f}\big)^2=g^{-1}f$, is crucial for the absence of the 
Boulware-Deser ghost \cite{Boulware:1973my}. We will briefly review
the ghost analysis of \eqref{action} in the next subsection.  

The bimetric action has a well-defined mass spectrum around
proportional backgrounds $\bar f_{\mu\nu}=c^2\bar g_{\mu\nu}$. 
Then the perturbations of the two metrics diagonalize into a massless
and a massive fluctuation, $\delta G\propto \delta g/m_f^2 + 
\delta f/m_g^2$ and $\delta M \propto \delta f- c^2 \delta g$
\cite{Hassan:2012wr}. Hence the metrics $\gmn$ and $\fmn$ are
combinations of massless and massive modes. The metric with a larger
Planck mass has a larger massless component.

The known ghost-free matter couplings that can be added to
\eqref{action} are of the form,  
\beqn
S_\mathrm{matter}=\int\dd^4x\sqrt{g}~\mathcal{L}_g
(g,\phi_g)+\int\dd^4x\sqrt{f}~\mathcal{L}_f (f,\phi_f)\,, 
\eeqn
where $\phi_g$ and $\phi_f$ denote different types of matter fields
that minimally couple to the respective metrics in the standard
way. In the limit $m_g>>m_f$, the metric $\gmn$ is mostly massless and
can be regarded as the gravitational metric with $M_p=m_g$
\cite{Hassan:2011zd,Hassan:2012wr}, while $\fmn$ is an extra spin-2
field that modifies gravity. In this framework, the observed high
scale $M_p$, or the weak strength, of gravity is correlated with the
effective masslessness of the interaction. 

Coupling both metrics to the same type of matter was considered in
\cite{Akrami:2013ffa} but it can be shown that such couplings reintroduce
the BD ghost \cite{Yamashita:2014fga, deRham:2014naa}. It is also
known that coupling the most obvious nonlinear extension of the
massless mode to matter also reintroduces the ghost
\cite{Hassan:2012wr}. 

\subsection{A new proposed matter coupling}\label{newcoupling}
Recently, \cite{deRham:2014naa} considered coupling matter to an
effective metric built out of $g$ and $f$ as,
\beqn
G_{\mu\nu}=a^2\gmn+2ab\,g_{\mu\rho}{S^\rho}_\nu+b^2\fmn\,,
\label{new}
\eeqn
where $a$ and $b$ are arbitrary parameters.\footnote{This combination
  can serve as a metric since $gS$ is symmetric, which can be easily
  seen by formally expanding $S=\sqrt{\mathbb 1 +(g^{-1}f-{\mathbb
      1})}$.} Note that for $a,b\neq 0$, these are degenerate with the
bimetric parameters and can always be set to one by rescalings
$g\rightarrow g/a^2$, $f\rightarrow f/b^2$ and absorbing the factors
that are generated in the bimetric action into $m_g$, $m_f$ and
$\beta_n$. Some interesting features of the new matter couplings were
discussed in \cite{deRham:2014naa}. Some other features will be
discussed below.

The coupling of $G$ to matter passes some simple necessary checks for
being ghost-free. \cite{deRham:2014naa} showed that in two simplifying
limits, the mini-superspace approximation and the decoupling limit,
the new couplings were ghost-free. But a more general Hamiltonian
analysis of perturbations around flat space showed a ghost at the
sixth order. Hence, \cite{deRham:2014naa} concludes that beyond the
approximations considered, the theory is not ghost-free.

Since the mass of the ghost seemed to be above the energy scale of the
theory, one may try to argue, as in \cite{deRham:2014naa}, that the
new matter couplings could still be considered in an effective theory
below a cutoff. Such an argument is obviously valid for non-ghost
excitations. But its validity for a ghost mode is in doubt since the
production of ghost modes do not require external energy, hence vacuum 
can spontaneously decay into ghost modes, releasing energy above the
cutoff scale. This would make an effective theory which contains a
ghost above the cutoff inherently inconsistent. For a discussion, see,
for example, \cite{Woodard:2006nt,Sbisa:2014pzo}. Hence the presence
of a ghost would be fatal for the new matter couplings irrespective of
its mass scale. 

The aim of this work is to perform a complete Hamiltonian analysis of
the new matter couplings and show that they do not reintroduce the
ghost into the bimetric theory.

\subsection{Review of ghost analysis in bimetric theory}\label{ADMbim}

Here we briefly review the main aspects of the ghost analysis of
bimetric theory
\cite{Hassan:2011hr,Hassan:2011tf,Hassan:2011zd,Hassan:2011ea}. This
also provides the framework for addressing the ghost issue in new
matter couplings. In particular, we emphasize a crucial aspect of the 
analysis not sufficiently clearly stated earlier.

To carry out a Hamiltonian analysis, we start with the usual ADM
decomposition \cite{Arnowitt:1962hi} of the two metrics, which in
matrix notation reads,
\beqn\label{admori}
g=
\begin{pmatrix}
-N^2+\nu^\mathrm{T}\gamma\nu~&~\nu^{\mathrm{T}}\gamma\\
\gamma\nu ~&~ \gamma
\end{pmatrix}\,, \qquad
f=
\begin{pmatrix}
-L^2+\lambda^\mathrm{T}\phi\lambda~&~\lambda^{\mathrm{T}}\phi\\
\phi\lambda ~&~ \phi
\end{pmatrix}\,.
\eeqn 
Here, $N$ and $L$ are the lapses, $\nu^i$ and $\lambda^i$ are the
shift vectors and $\gamma_{ij}$ and $\phi_{ij}$ are the spatial
3-metrics. In these variables, the bimetric Lagrangian takes the form
(in phase space variables and up to surface terms),  
\begin{align}
\label{action2}
\mathcal{L}= 
\pi^{ij}\partial_t\gamma_{ij}+p^{ij}\partial_t\phi_{ij}
+N R^{0(g)}+\nu^iR_i^{(g)}+L R^{0(f)}+\lambda^iR_i^{(f)}
-2m^4\tilde V(N,L,\nu,\lambda,\gamma,\phi)\,.
\end{align}
Here $\pi^{ij}$ and $p^{ij}$ are the momenta canonically conjugate to
$\gamma_{ij}$ and $\phi_{ij}$. The 8 lapses and shifts appear without
time derivatives and are non-dynamical, while, a priori, $\gamma_{ij}$
and $\phi_{ij}$ contain 12 dynamical fields (24 phase space degrees of
freedom). These include ghost modes that must be removable by gauge
fixing and by constraints arising from the equations of motion for
$N,L,\nu^i,\lambda^i$. But since $\tilde V$ is highly nonlinear in the
$N,L,\nu^i,\lambda^i$, the corresponding 8 equations of motion could
potentially depend on all the non-dynamical variables and determine
them in terms of $\gamma_{ij}$ and $\phi_{ij}$, rather than becoming
constraints on $\gamma_{ij}$ and $\phi_{ij}$. This is the basis of the
Boulware-Deser argument for presence of a ghost in such theories
\cite{Boulware:1973my}. However, it is also possible that these
equations do not determine all lapses and shifts in which case some of
them will instead impose constraints on $\gamma_{ij}$ and
$\phi_{ij}$.\footnote{In the massive gravity context this possibility
  was discussed in \cite{deRham:2010ik, deRham:2010kj}} We outline the analysis below.   

If it turns out that the 8 equations of motion arising from
$N,L,\nu^i,\lambda^i$ can determine only 3 combinations of these
variables, say, $n^i(N,L,\nu,\lambda,\gamma,\phi)$, in terms of
$\gamma_{ij}$ and $\phi_{ij}$, then the remaining 5 equations will not
depend on the non-dynamical variables. They become constraints on
$\gamma_{ij}$ and $\phi_{ij}$ and eliminate some of the unwanted
modes.\footnote{This counting is for ghost-free bimetric theory
\cite{Hassan:2011zd}. Other counting possibilities have been
considered in \cite{Comelli:2012vz,Comelli:2014xga} in the massive
gravity context.}    

In the action let us now trade 3 of the non-dynamical variables, say,
the $\nu^i$, for the combinations $n^i$ (the possibilities are
restricted by 3-dimensional general covariance). For the above
picture to hold, the remaining 5 variables $\lambda^i$, $N$ and $L$
must now appear linearly as Lagrange multipliers that enforce the 5
constraint equations. Since the action \eqref{action2} already
contains a term $\nu^iR_i^{(g)}$, this in turn implies that the
expression for $\nu^i$ in terms of $n^i$ must be linear in the
remaining lapses and shifts. For the bimetric action \eqref{action}, 
a combination that works is \cite{Hassan:2011tf,Hassan:2011zd},    
\beqn
\label{redef}
\nu^i=\lambda^i+ Ln^i+ N {D^i}_jn^j\,.
\eeqn 
The metric $D^i_{~j}$ will be specified below. On eliminating $\nu^i$
in favor of $n^i$, the action takes the form \cite{Hassan:2011zd}, 
\beqn\label{actionADM}
S=\int\dd^4x\left(\pi^{ij}\p_t\gamma_{ij}+p^{ij}\p_t\phi_{ij} 
+\lambda^i\mathcal{C}^{(\lambda)}_i+L\mathcal{C}_L+N\mathcal{C}_N
\right)\,, 
\eeqn
where $\mathcal{C}_i^{(\lambda)}=R_i^{(g)}+R_i^{(f)}$ is independent of
$n^i$. The explicit forms of $\mathcal{C}_N$ and $\mathcal{C}_L$ are
given below. The important point is that in terms of $n^i$, the theory
is linear in the $\lambda^i$, $L$ and $N$. 

Since the $n^i$ equations of motion are linear in $L$ and $N$,
naively it seems that $n^i$ depend on the lapses, contrary to the 
assumption. However, since $n^i$ enter only through
$\nu^i$, we have,
\beqn
\frac{\delta S}{\delta n^i}=
\left(\frac{\delta S}{\delta \nu^j}\right)_{\nu=\nu(n)}
\frac{\delta\nu^j}{\delta n^i} =0\,,
\eeqn 
or explicitly,
\beqn
N\frac{\p}{\p n^i}{\cal C}_N +L\frac{\p}{\p n^i}{\cal C}_L
={\cal C}^{(\nu)}_j\left(L\delta^j_{~}i+N\frac{\p}{\p n^j}
(Dn)^i\right)=0\,.
\label{ni-preeom}
\eeqn 
Note that both sides of the first equality are linear in $N$ and
$L$, hence it is obvious that ${\cal C}^{(\nu)}_j$, when expressed as 
a function of $n^i$, must be independent of the lapses. Since the
Jacobian factor ${\delta\nu^j}/{\delta n^i}$ is invertible, it follows
that $n^i$ are determined by the equations  
\beqn
{\cal C}^{(\nu)}_j(n,\gamma,\phi,\pi,p)=0\,,
\eeqn
and are independent of the lapses and shifts as desired. This property
has been explicitly verified for the action \eqref{action}
\cite{Hassan:2011hr,Hassan:2011tf,Hassan:2011zd,Hassan:2011ea},  
but the above discussion shows that this is always the case whenever
the action can be made linear in the lapses and shifts 
through field redefinitions of the type \eqref{redef}. The Lagrange
multipliers in \eqref{actionADM} now lead to 5 constraints on the
a priori dynamical variables,
\beqn
\mathcal{C}^{(\lambda)}_i=0\,,\qquad
\mathcal{C}_L=0\,,\qquad
\mathcal{C}_N=0\,.
\eeqn
 The last of these is accompanied by a secondary constraint
 $\mathcal{C}_{N(2)}\equiv\p_t\mathcal{C}_N=0$ and the pair remove
 the Boulware-Deser ghost and its conjugate momentum
 \cite{Hassan:2011ea}. The other 4 constraints are associated with
 general covariance and along with gauge fixing, eliminate another 8
 phases pace degrees of freedom, reducing to phase space degrees of
 freedom to 24-2-8=14, or 7 dynamical fields. Some of the original
 dynamical equations now reduce to further constraints that determine 
 $N$, $L$ and $\lambda^i$. As an aside, \eqref{ni-preeom} implies that
 when the $n^i$ equations are satisfied, then $\p{\cal C}_N/{\p
   n^i}=0$ and $\p{\cal C}_L/{\p n^i}=0$, hence all constraints become
 $n^i$-independent.  

The feasibility of proving the absence of ghost in this way depends on
the possibility of converting an action that is non-linear 
in the lapses and shifts \eqref{action}, to a partially linear form
\eqref{actionADM} through a redefinition of the form \eqref{redef}. In
the bimetric case, a further complication is the appearance of the
square-root matrix in the action. The redefinition \eqref{redef}
resolves both these problems provided the 3$\times$3 matrix $D^i_{~j}$
in \eqref{redef} satisfies a condition that solves to
\cite{Hassan:2011tf},  
\beqn\label{dx}
D=\sqrt{\gamma^{-1}\phi(x\mathbb{1}+nn^\mathrm{T}\phi)}~(x\mathbb{1}
+nn^\mathrm{T}\phi)^{-1}\,, \qquad
x\equiv 1-n^\mathrm{T}\phi n\,.
\eeqn
Here we use matrix notation and $\mathbb{1}$ stands for the matrix
with components $\delta^i_j$. In terms of the $n^i$, the action
takes the form \eqref{actionADM} where $\mathcal{C}_L$ and
$\mathcal{C}_N$ are given by \cite{Hassan:2011zd},  
\begin{align}
\mathcal{C}_L&=R_0^{(f)}+R_i^{(g)}n^i +2m^4 \sqrt{\det\gamma}\,U \,,\nn\\
\mathcal{C}_N&=R_0^{(g)}+R_i^{(g)}{D^i}_jn^j +2m4 \sqrt{\det\gamma}\,V\,.
\end{align}
$U$ and $V$ are the contributions from the interaction potential
and read, in matrix notation,  
\begin{align}\label{UV}
U=&\sqrt{x}\left(\sum_{n=0}^2\beta_{n+1}e_n(\sqrt{x}D)+\beta_3\left(e_1(D)\,n^\mathrm{T}\phi Dn-(Dn)^\mathrm{T}\phi Dn\right)\right)+\beta_2n^\mathrm{T}\phi Dn+\beta_4\frac{\sqrt{\det \gamma}}{\sqrt{\det \phi}}\,,\nn\\
V=&\sum_{n=0}^3\beta_ne_n(\sqrt{x}\,D)\,.
\end{align}
\section{A new redefinition of shift variables}\label{adm2}

The present form of the redefinition (\ref{redef}) is asymmetric in
the Hamiltonian variables for the two metrics. In this section we
introduce a new parametrization of the shift vectors $\nu^i$ which
appears more symmetric. The new variables simplify the expressions and
facilitate the ghost analysis for the new matter coupling. Later, in 
section \ref{symvielbein}, we show that the new shift variables are 
related to the Lorentz boost that symmetrizes a combination of the
vielbeins for $g$ and $f$ such that the square root $\sqrt{g^{-1}f}$
can be evaluated.  

\subsection{The new redefinition}

We begin by decomposing the spatial metrics into vielbeins, 
\beqn
\gamma_{ij}={e^a}_i\delta_{ab}{e^b}_j \,,\qquad
\phi_{ij}={\varphi^a}_i\delta_{ab}{\varphi^b}_j\,.
\eeqn
The vielbeins are defined up to two independent local Lorentz
rotations. Let us consider the following further redefinition of the
shift vector $n^i$ in terms of new variables $v_a$,  
\beqn\label{nnew}
n^i={(\tilde\varphi^{-1})^i}_a\delta^{ab}v_b\,.
\eeqn
Here, $\tilde\varphi=R\varphi$ and $R$ is a specific Lorentz rotation
obtained below. Next, we express the matrix $D$ in (\ref{dx}) in terms
of the new variables,  
\beqn\label{D2}
D=\sqrt{e^{-1}\JJ^{-1}(e^{-1})^\mT \varphi^\mT R^\mT
(x\JJ+vv^\mT)R\varphi}~\big(x\mathbb{1}+\varphi^{-1}R^{-1}\JJ^{-1}
vv^\mT R\varphi\big)^{-1}\,,
\eeqn
where now $x=1-v^\mT\JJ^{-1}v$ and for the matrices that raise and
lower local Lorentz indices we have introduced the following symbols,  
\beqn\label{idef}
\JJ_{ab}=\delta_{ab}\,,\qquad
(\JJ^{-1})^{ab}=\delta^{ab}\,.
\eeqn
Observe that we can write,
\beqn
x\JJ+vv^\mT=x(\JJ+\tfrac{1}{x+\sqrt{x}}vv^\mT)\JJ^{-1}(\JJ+
\tfrac{1}{x+\sqrt{x}}vv^\mT)\,. 
\eeqn
The square root of the 3$\times$3 matrix in (\ref{D2}) can be
evaluated if we demand the following symmetry property, 
\beqn\label{symcond}
\Big[\left( \JJ+\tfrac{1}{x+\sqrt{x}}vv^\mT\right)R\varphi
  e^{-1}\Big]^\mT =\left(\JJ+\tfrac{1}{x+\sqrt{x}}vv^\mT\right)
R\varphi e^{-1}\,. 
\eeqn
This requirement can always be satisfied by choosing an appropriate
local Lorentz rotation $R$. The quantity considered is of the form
$ARB$ with $R^\mT\JJ R=\JJ$ and can be symmetrized by,
\beqn
R= \sqrt{(A^\mT B^{-1}\JJ^{-1})(A^\mT B^{-1}\JJ^{-1})^\mT}
\,(A^\mT B^{-1}\JJ^{-1})^\mT.
\eeqn
Here $A=A^\mT=(\JJ+\tfrac{1}{x+\sqrt{x}}vv^\mT)$ and $B=\varphi
e^{-1}$. This is a sandwiched version of the standard polar
decomposition, the matrix under the square-root is now positive and
$R$ always exists, exactly as in polar decomposition (see appendix
\ref{sandwich} for a derivation).    

Now, using the notation $\tilde\varphi=R\varphi$, with $R$ fixed as
above, the matrix $D$ becomes, 
\begin{align}
\label{Dnew1}
D&=\sqrt{x}\sqrt{e^{-1}\JJ^{-1}(e^{-1})^\mT\tilde\varphi^\mT
\left( \JJ+\tfrac{1}{x+\sqrt{x}}vv^\mT\right)\JJ^{-1}
\left( \JJ+\tfrac{1}{x+\sqrt{x}}vv^\mT\right)\tilde\varphi}~
\big(x\mathbb{1}+\tilde\varphi^{-1}\JJ^{-1}vv^\mT\tilde\varphi\big)^{-1}
\nn\\
&=\sqrt{x}\sqrt{e^{-1}\JJ^{-1}\left(\JJ+\tfrac{1}{x+\sqrt{x}}vv^\mT\right) 
\tilde\varphi e^{-1}\JJ^{-1}\left(\JJ+\tfrac{1}{x+\sqrt{x}}vv^\mT\right)
\tilde\varphi}~\big(x\mathbb{1}+\tilde\varphi^{-1}\JJ^{-1}vv^\mT
\tilde\varphi\big)^{-1}\nn\\
&=\sqrt{x}~e^{-1}\JJ^{-1}\left( \mathbb{1}+
\tfrac{1}{x+\sqrt{x}}vv^\mT\right)\tilde\varphi\big(x\mathbb{1}
+\tilde\varphi^{-1}\JJ^{-1}vv^\mT\tilde\varphi\big)^{-1}\,.
\end{align}
Finally, using,
\beqn
(x\mathbb{1}+\tilde\varphi^{-1}\JJ^{-1}vv^\mT\tilde\varphi)^{-1}
=x^{-1}(\mathbb{1}-\tilde\varphi^{-1}\JJ^{-1}vv^\mT\tilde\varphi)\,,
\eeqn
we arrive at the simple expression,
\beqn\label{Dnew}
D=\tfrac{1}{\sqrt{x}}\,e^{-1}\left(\hat{\mathbb{1}}-
\tfrac{1}{1+\sqrt{x}}\JJ^{-1}vv^\mT\right)\tilde\varphi\,,
\eeqn
where $\hat{\mathbb{1}}^a_b=\delta^a_b$. When acting on
$n^i={(\tilde\varphi^{-1})^i}_a\delta^{ab}v_b$, this matrix gives
${D^i}_jn^j={(e^{-1})^i}_{a}\delta^{ab}v_b$. Putting everything
together, we find that the redefinition (\ref{redef}) of the original 
shift vector now reads, 
\beqn\label{redefnew}
\nu^i=\lambda^i+\Big(L{(\tilde\varphi^{-1})^i}_{a}
+N{(e^{-1})^i}_{a}\Big)\delta^{ab}v_b\,,
\eeqn
which has a simpler symmetric form. But note that the gauge fixed
$\tilde\varphi^a_{~i}$ now also depends on $\phi^a_{~i}$, $e^a_{~i}$
and $v_a$ through $R^a_{~b}$. This complication however does not
affect the ghost argument.

The new variables also answer another question: The expression
\eqref{dx} for $D$, which appeared in the massive gravity and bimetric
ghost analysis, involves a $3\times 3$ square-root matrix the
existence of which is not evident. The analysis here shows that in the 
Hamiltonian framework employed, the matrix $D$ always exists.

The action in terms of the new variables can easily be obtained
from~(\ref{actionADM}) where one replaces the spatial metrics in terms
of vielbeins as well as $n^i$ and ${D^i}_j$ in terms of $v_a$
using~(\ref{nnew}) and~(\ref{Dnew}). Clearly, with the new
redefinition, the ghost proof of~\cite{Hassan:2011tf, Hassan:2011zd,
  Hassan:2011ea} which we reviewed in section~\ref{ADMbim} goes
through in the same way as before, with $n^i$ replaced by $v_a$.

\section{Absence of ghost in the new matter coupling}\label{consproof}

We now turn to the ghost analysis of the matter coupling for a
composite metric proposed in \cite{deRham:2014naa}, where it was also
concluded that the new couplings reintroduced the Boulware-Deser ghost
instability into the theory. This would render the new couplings
unusable even in an effective theory sense, as discussed in section
\ref{newcoupling}. However, using the new bimetric variables $v^a $ ,
we show that the new matter couplings are linear in the lapses $N$,
$L$ and the shift $\lambda^i$. Moreover, the equations for $v_a$ are
independent of $N$, $L$ and $\lambda^i$. Thus, the theory, including
the new matter coupling, contains the same number of constraints as
pure bimetric theory and hence should be free of the Boulware-Deser
ghost mode. 

\subsection{Matter coupling of the effective metric}

Recently, \cite{deRham:2014naa} proposed coupling the ghost-free
bimetric theory to matter through an ``effective'' metric, 
\beqn\label{effm}
G_{\mu\nu}=a^2\gmn+2ab\,g_{\mu\rho}{\left(\sqrt{g^{-1}f}\right)^\rho}_\nu
+b^2\fmn\,.
\eeqn
As argued in section \ref{newcoupling}, we can set $a=b=1$ without
loss of generality. Then, $G=g(1+S)^2$. 

Let us denote the ADM variables of $G$ by $N_\mathrm{eff}$
(lapse), $\nu^i_\mathrm{eff}$ (shift) and $(\gamma_\mathrm{eff})_{ij}$
(spatial metric), such that, 
\beqn\label{geffadm}
G=\begin{pmatrix}
-N_\mathrm{eff}^2+\nu_\mathrm{eff}^k(\gamma_\mathrm{eff})_{kl}\nu_\mathrm{eff}^l 
 ~~&~~  \nu_\mathrm{eff}^k(\gamma_\mathrm{eff})_{kj}\\
(\gamma_\mathrm{eff})_{ik}\nu_\mathrm{eff}^k  ~~&~~(\gamma_\mathrm{eff})_{ij}
\end{pmatrix}\,. 
\eeqn
From general relativity it is known that, for standard minimal matter
couplings, the matter Lagrangian expressed in terms of phase space
variables, takes the form,\footnote{For coupling to fermions, one
  needs to invoke the vierbein (\ref{sumvier}) for $G$. Also in this
  case, it follows from standard results in general relativity that
  the vierbein-matter interactions are linear in $N_\mathrm{eff}$ and 
  $\nu^i_\mathrm{eff}$.} 
\beqn\label{mattadm}
\mathcal{L}_\mathrm{matter}=\mathcal{L}_0+N_\mathrm{eff}
\Theta+\nu^i_\mathrm{eff} \Theta_i\,. 
\eeqn
This is linear in $N_\mathrm{eff}$ and $\nu^i_\mathrm{eff}$, which,
however, are highly nonlinear in the ADM variables of $g$ and 
$f$. In order not to spoil the consistency of the bimetric potential,
$\mathcal{L}_\mathrm{matter}$ must become linear in $N$, $L$, and
$\lambda^i$ after the redefinition \eqref{redefnew} has been
performed. We thus need to show that $N_\mathrm{eff}$ and
$\nu_\mathrm{eff}^i$ are linear functions of $N$, $L$, and $\lambda^i$ 
and that the latter do not appear in the $v_a$ equations of motion. 

\subsection{Linearity in the lapses}
It is straightforward to work out the expressions for
$\nu_\mathrm{eff}$ and $\gamma_\mathrm{eff}$. First, we note that
after the redefinition (\ref{redefnew}) the ADM decomposition of the 
term $g\sqrt{g^{-1}f}$ in $G$ reads, 
\begin{align}
&g\sqrt{g^{-1}f}=\nn\\
&\quad\qquad\begin{pmatrix}
-\sqrt{x} NL + (\lambda+L\tilde\varphi^{-1}\JJ^{-1}v)^\mT e^\mT\chi
\tilde\varphi(\lambda+L\tilde\varphi^{-1}\JJ^{-1}v)
~&~(\lambda+L\tilde\varphi^{-1}\JJ^{-1}v)^\mT e^\mT\chi\tilde\varphi\\
 e^\mT\chi\tilde\varphi(\lambda+L\tilde\varphi^{-1}\JJ^{-1}v)~&~ 
e^\mT\chi\tilde\varphi
\end{pmatrix}.
\end{align}
To keep the expression shorter, we have used matrix notation and
defined,
\beqn\label{chidef}
\chi_{bc}=\delta_{bc}-\tfrac{1}{1+\sqrt{x}}v_bv_c\,.
\eeqn
Plugging this together with the ADM decompositions for $g$ and $f$
into the effective metric $G=g+2gS +f$ and comparing the result
to~(\ref{geffadm}), we can easily read off the spatial metric and the
shift of $G$,\footnote{Note that the symmetry of the matrix
  $e^\mT\chi\tilde\varphi$ is equivalent to the symmetry of $(
  \JJ+\tfrac{1}{x+\sqrt{x}}vv^\mT)\tilde\varphi e^{-1}$ which is
  imposed by our choice of rotational gauge in~(\ref{symcond}).} 
\beqn
(\gamma_\mathrm{eff})_{ij}&=&\gamma_{ij}+\phi_{ij}+{e^a}_i\chi_{ab}
{\tilde\varphi^b}_j+{e^a}_j\chi_{ab}{\tilde\varphi^b}_i\,,\\
\nu_\mathrm{eff}^i&=& \lambda^i+L {(\tilde\varphi^{-1})^i}_a
\delta^{ab}v_b+(\gamma_\mathrm{eff}^{-1})^{ij}(N{e^a}_j-
L{\tilde\varphi^a}_j)v_a\,.
\eeqn

In principle, the lapse $N_\mathrm{eff}$ can be derived in the same
manner, but this requires a tedious computation (which we have
performed, verifying that the result agrees with the expressions 
derived below). For the sake of transparency, we provide a 
simpler derivation here. A third approach is presented in
appendix \ref{altder}. Consider,
\beqn
\sqrt{\det g_{\mathrm{eff}}}=N_\mathrm{eff}
\sqrt{\det\gamma_{\mathrm{eff}}}\,.  
\eeqn
Using \eqref{det-e}, this is also equal to,
\beqn\label{detg}
\sqrt{\det g_{\mathrm{eff}}}= \sqrt{\det g}\,\det\left(\mathbb{1}
+\sqrt{g^{-1}f}\right) =N\sqrt{\det \gamma}\,\sum_{n=0}^4 
e_n\left(\sqrt{g^{-1}f}\right)\,.
\eeqn
Thus we have,
\beqn\label{efflap}
N_\mathrm{eff}=
\frac{1}{\sqrt{\det \gamma_{\mathrm{eff}}}}\left(N\sqrt{\det \gamma}
\sum_{n=0}^4 e_n\left(\sqrt{g^{-1}f}\right)\right)\,.
\eeqn
Note that the right-hand side of (\ref{detg}) is the bimetric
potential with all $\beta_n$ set to one. After the 
redefinition of the shift $\nu^i$, this expression is linear in the
$N$ and $L$ and does not contain the shifts $\lambda^i$, as discussed
in section \ref{ADMbim}. More precisely, we have,  
\beqn
N\sqrt{\det \gamma}\sum_{n=0}^4 e_n(S)=N\sqrt{\det \gamma}\,\,
V+L\sqrt{\det \gamma}\,\, U\,, 
\eeqn
where the scalar functions $U$ and $V$ are defined as in (\ref{UV})
but with $\beta_k=1$ for all $k$. In terms our new shift variables
$v_a$ \eqref{redefnew} they read,
\begin{align}
U&=\sqrt{x}\left(\sum_{n=0}^2e_n(\sqrt{x}\,D)+e_1(D)\,v^\mT
\tilde\varphi e^{-1}\JJ^{-1} v-(\tilde\varphi e^{-1}\JJ^{-1} v)^\mT\JJ 
\tilde\varphi e^{-1}\JJ^{-1} v\right)\nn\\
&~\hspace{225pt} +~v^\mT\tilde\varphi e^{-1}\JJ^{-1} v+
\det(\tilde\varphi e^{-1})\,,\nn\\
V&=\sum_{n=0}^3e_n(\sqrt{x}\,D)\,,
\end{align}
where,
\beqn
D=\tfrac{1}{\sqrt{x}}e^{-1}\left(\hat{\mathbb{1}}-\tfrac{1}{1+\sqrt{x}}
\JJ^{-1}vv^\mT\right)\tilde\varphi\,.
\eeqn
This shows that the effective lapse $N_\mathrm{eff}$ and shift
$\nu_\mathrm{eff}^i$ are linear functions of $N$, $L$ and $\lambda^i$.
Hence, the matter coupling will not introduce nonlinearities for these
variables.   

\subsection{Absence of the lapses in the shift equations}\label{shift}

Let us recapitulate what we have established so far. After the
redefinition that renders the bimetric potential linear in $N$, $L$
and $\lambda^i$, the same holds for the matter coupling. Thus the
complete bimetric plus matter Lagrangian takes the form,  
\beqn
\mathcal{L}=\mathcal{L}_\mathrm{dyn}'+\lambda^i\mathcal{C}'^{(\lambda)}_i
-N\mathcal{C}'_N-L\mathcal{C}'_L\,, 
\eeqn  
where $\mathcal{L}_\mathrm{dyn}'$ contains the kinetic terms
in~(\ref{actionADM}) plus any new dynamics from the matter sector,
i.e. the terms in \eqref{mattadm}) that do not depend on $N$, $L$
and~$\lambda^i$. The constraints $\mathcal{C}'^{(\lambda)}_i$,
$\mathcal{C}'_N$ and $\mathcal{C}'_L$ consist of the original terms
from the bimetric potential as well as the new contributions from the
matter coupling~(\ref{mattadm}). Variation of the Lagrangian with
respect to $N$, $L$ and~$\lambda^i$ produces equations that are
independent of the variables themselves and hence constrain the
remaining phase space degrees of freedom. We note further that
$\mathcal{C}'^{(\lambda)}_i$, does not depend on the components of
$v_a$ and consequently $\lambda^i$ does not show up in their
equations. 

However, $\mathcal{C}'_N$ and $\mathcal{C}'_L$ depend on the $v_a$ and
the only thing left to check is that $N$ and $L$ do not appear in the
equations of motion for $v_a$. This is crucial because if the $v_a$
equations did depend on $N$ or $L$, then they would not determine
$v_a$ in terms of the phase space variables alone. Consequently there
would not exist enough constraints on the latter to eliminate the
ghost.\footnote{One can also see the absence of the lapse constraint
  in the case where the shift equations depend on the lapse as
  follows: If the solution for the shift depends on the lapse, then
  nonlinear functions of the lapse will appear in the action after the
  shift has been integrated out. Hence, the lapse equation of motion
  will now determine the lapse itself instead of imposing a constraint
  on other variables.} In this case, the Boulware-Deser ghost would
propagate and destroy the consistency of the theory. 

We thus need to demonstrate that the~$v_a$ equations do not determine
$N$ nor $L$. The argument for this is the same as that outline for the
pure bimetric theory. Naively, one finds that $N$ and $L$ multiply
nonlinear functions of $v_a$ and are thus expected to appear in its
equations. However, the new shift vector $v_a$ enters the action only
through the original variable $\nu^i$ \eqref{redefnew}. Hence its
equations of motion can be written as, 
\beqn
\label{v-eom}
0=\frac{\delta S}{\delta v_a}=\frac{\delta S}{\delta \nu^j}
\frac{\delta \nu^j}{\delta v_a}=\frac{\delta S}{\delta \nu^j}
\left(N(e^{-1})^{ja}+L \frac{\delta}{\delta v_a}
\big({(\tilde\varphi^{-1})^j}_b\delta^{bc}v_c\big)\right)\,.
\eeqn
Now, since the action is linear in the lapses, so is the variation
with respect to $v^a$ on the left-hand-side. Since the Jacobian factor
$\frac{\delta \nu^j}{\delta v_a}$ is already linear in the lapses, it
follows that $\frac{\delta S}{\delta \nu^j}$, when expressed in terms
of the $v_a$, cannot depend on $L$ and $N$ (otherwise there would be
nonlinear terms). Furthermore, in order for the redefinition to be
well defined, we need the Jacobian to be invertible and hence, the
$v_a$ equations of motion are equivalent to $\frac{\delta S}{\delta
  \nu^j}=0$, which does not involve the lapses. We emphasize the
generality of this statement: For a redefinition that renders the
Lagrangian linear in the lapses and that is linear in the lapses
itself, the equation for the redefined shift vector are always
independent the the lapses.

This completes the proof that the number of constraints in bimetric
theory is not altered when the composite metric (\ref{effm}) is coupled  
to matter and hence the Boulware-Deser ghost is not reintroduced by
the novel matter coupling.

Our result is at variance with the conclusion in \cite{deRham:2014naa}
that in the presence of the new matter couplings, the theory is no
longer linear in the lapses and hence is not ghost-free. Since
\cite{deRham:2014naa} provides the result of a perturbative analysis
without the calculational details, the discrepancy remains
unexplained. 

\section{Relation to vielbein formulation} \label{symvielbein}

This section is devoted to the interesting interpretation of the new
bimetric variables introduced in section \ref{adm2} in the context of
the vierbein formulation of bimetric theory. The consistency proof of
the matter coupling in the previous section does not rely on this
background material. 

\subsection{Symmetrization condition}

A reformulation of bimetric theory in terms of vierbeins which avoids
the square-root matrix has been proposed in
\cite{Hinterbichler:2012cn}. In order to express the interaction
potential of the metric formulation in terms of vierbeins, 
we decompose the two metrics, $\gmn={e(g)^a}_\mu\eta_{ab}{e(g)^b}_\nu$
and $\fmn={e(f)^a}_\mu\eta_{ab}{e(f)^b}_\nu$. The square root
$\sqrt{g^{-1}f}$ can be evaluated in terms of the vierbeins provided
that the following symmetry condition is satisfied, 
\beqn\label{symcondv}
\eta_{ac}\, {e(f)^c}_\mu {\big(e(g)^{-1}\big)^\mu}_b=\eta_{bc}\, {e(f)^c}_\mu {\big(e(g)^{-1}\big)^\mu}_a\,.
\eeqn
If this is the case one obtains, in matrix notation,
\beqn
 e(g)^{-1} e(f)e(g)^{-1} e(f)=e(g)^{-1}\eta^{-1}(e(g)^{-1})^\mT e(f)^\mT\eta e(f)=g^{-1}f\,,
\eeqn
and hence,
\beqn
 \sqrt{g^{-1}f}=e(g)^{-1}e(f)\,.
\eeqn
The question whether the symmetry condition (\ref{symcondv}) follows
from the vierbein equations or can be implemented in the dynamics by
extending the theory is beyond the scope of this paper. For related
work, see \cite{Hassan:2012wt, Deffayet:2012zc, Banados:2013fda}. 

Here we are interested in exploring the relation between the symmetry
condition and our new redefinition of the shift vector. A general
vierbein can be parametrized by a Lorentz boost acting on a
triangular vierbein. We therefore start with partially gauge fixed
vierbeins in triangular form, 
\beqn
e_\mathrm{t}(g)= \begin{pmatrix}
N~&~0\\
{e^a}_k\nu^k ~&~ {e^a}_i
\end{pmatrix}
\,,\qquad
e_\mathrm{t}(f)=\begin{pmatrix}
L~&~0\\
{\varphi^a}_k\lambda^k~&~ {\varphi^a}_i
\end{pmatrix}\,,
\eeqn 
whose components translate into the ADM variables~(\ref{admori}) of
the metrics with $\gamma_{ij}={e^a}_i\delta_{ab}{e^b}_j$ and
$\phi_{ij}={\varphi^a}_i\delta_{ab}{\varphi^b}_j$. 
In the above Lorentz gauge, the combined matrix $\hat{S}\equiv
e_\mathrm{t}(f)e_\mathrm{t}(g)^{-1}$ reads, 
\beqn\label{sdef}
\hat{S}\equiv\begin{pmatrix}
\Sigma~&~0\\
\sigma^a ~&~ {s^a}_b
\end{pmatrix}=
\begin{pmatrix}
\tfrac{L}{N}~&~0\\
\tfrac{1}{N}{\varphi^a}_k(\lambda^k-\nu^k)~&~ {\varphi^a}_k {(e^{-1})^k}_b
\end{pmatrix}\,.
\eeqn
Note that for the triangular vierbeins the symmetry
condition~(\ref{symcondv}) is not satisfied by~$\hat{S}$. 
However, the general form for $\hat{S}$ is given by a Lorentz boost
acting on~(\ref{sdef}). In order to make the connection to the metric
formulation, we thus need to find a Lorentz boost $\Lambda$ that
symmetrizes~$\hat{S}$, 
\beqn
S_\mathrm{s}=\eta\Lambda \hat{S}\,,\qquad S_\mathrm{s}^\mT=S_\mathrm{s}\,.
\eeqn
The rotational subgroup of the Lorentz transformations is not fixed by
the triangular form. To account for the rotations we could write
$\varphi=R\varphi'$, where $\varphi'$ is the gauge fixed spatial
metric. For notational simplicity we refrain from doing so here but
keep in mind that $\varphi$ still contains the rotational degrees of
freedom. 

We now derive the expression for the Lorentz boost that achieves the
symmetrization of $\hat{S}$. A general boost can be parametrized in
terms of a boost velocity vector $v_a$ and the corresponding boost
factor $\Gamma=\big(\sqrt{1-v_a\delta^{ab}v_b}\,\big)^{-1}$, 
\beqn\label{LT}
\Lambda=\begin{pmatrix}
\Gamma~~&~~\Gamma  v_b\\
\Gamma v^a ~~&~~ \delta^a_b+\tfrac{\Gamma^2}{1+\Gamma}v^av_b
\end{pmatrix}\,.
\eeqn
Here and in what follows, $v^a$ with upper indices denotes $\delta^{ab}v_b$.
For the general form of $S_\mathrm{s}$ we now have,
\beqn
S_\mathrm{s}=\eta\Lambda \hat{S}=\begin{pmatrix}
-\Gamma \Sigma - \Gamma v^c\sigma_c ~~&~~ -\Gamma v_c {s^c}_b \\
\Gamma\Sigma  v_a+ \left( \delta_{ac}+\tfrac{\Gamma^2}{1+\Gamma} v_av_c\right)\sigma^c ~~&~~ \left( \delta_{ac}+\tfrac{\Gamma^2}{1+\Gamma}v_av_c\right){s^c}_b
\end{pmatrix}\,.
\eeqn
Demanding its symmetry gives the following conditions,
\begin{subequations}
\beqn
-\Gamma v_c{s^c}_a&=&\Gamma\Sigma v_a+ \left(  \delta_{ac}+\tfrac{\Gamma^2}{1+\Gamma}v_av_c\right)\sigma^c \,,\label{cond1}\\
\left( \delta_{ac}+\tfrac{\Gamma^2}{1+\Gamma}v_av_c\right){s^c}_b&=&\left( \delta_{bc}+\tfrac{\Gamma^2}{1+\Gamma}v_bv_c\right){s^c}_a\,.\label{cond2}
\eeqn
\end{subequations}
The second condition can be met by fixing the rotations in
$\varphi=R\varphi'$. As we argued before, this is always possible
thanks to the polar decomposition theorem. We therefore focus on the
first condition that we try to satisfy by fixing the velocity
vector~$v_a$ of the Lorentz boost. 

Multiplying the first condition~(\ref{cond1}) with $v_b$ results in
the following matrix equation, 
\beqn
-\Gamma v_bv_c{s^c}_a&=&\Gamma\Sigma v_bv_a + v_b\left( \delta_{ac}+\tfrac{\Gamma^2}{1+\Gamma}v_av_c\right)\sigma^c  \,.
\eeqn
Subtracting its transpose from the equation gives,
\beqn
\Gamma(v_av_c{s^c}_b-v_bv_c{s^c}_a)&=& v_b\sigma^c\delta_{ca}-v_a\sigma^c\delta_{cb} \,,
\eeqn
which we insert into the second condition~(\ref{cond2}) to get,
\beqn
\delta_{ac}{s^c}_b-\delta_{bc}{s^c}_a&=&\tfrac{\Gamma}{1+\Gamma}(  v_a\sigma^c\delta_{cb}-v_b\sigma^c\delta_{ca})\,.
\eeqn
Contraction with $v^b$ leads to,
\beqn
\left(\delta_{ac}{s^c}_b-\delta_{bc}{s^c}_a\right)v^b&=&\tfrac{\Gamma}{1+\Gamma}(  v_a\sigma^c\delta_{cb}-v_b\sigma^c\delta_{ca})v^b \nn\\
&=&\tfrac{\Gamma}{1+\Gamma}(\Gamma^{-2}-1) \sigma^c\delta_{ca} +\tfrac{\Gamma}{1+\Gamma}v_a\sigma^bv_b\,,
\eeqn
which implies,
\beqn
\left( \delta_{ab}+\tfrac{\Gamma^2}{1+\Gamma}v_av_b\right)\sigma^b=\left(\delta_{ac}{s^c}_b-\delta_{bc}{s^c}_a\right)v^b\,.
\eeqn
Plugging this back into the first condition~(\ref{cond1}), we arrive at,
\beqn
({s^a}_b+\Sigma \delta^a_b)\,v^b=-\sigma^a\,,
\eeqn
which finally yields the solution for the velocity vector of the Lorentz boost that symmetrizes the matrix $\hat{S}=e(f)e(g)^{-1}$, 
\beqn\label{vsol}
v^a=-{\big[(s+\Sigma \hat{\mathbb{1}})^{-1}\big]^a}_b~\sigma^b\,.
\eeqn
In terms of the components of $e(g)$ and $e(f)$ this expression reads,
\beqn\label{vsol2}
v^a={\big[(Ne^{-1}+L\varphi^{-1})^{-1}\big]^a}_i~(\nu^i-\lambda^i)\,.
\eeqn
Remarkably, this equation is exactly the same as (\ref{redefnew})
which means that the redefined shift vector of our new ADM variables
can be identified with the velocity vector of the Lorentz boost that
symmetrizes the matrix $\hat{S}$. Note also that the scalar $x$ in
(\ref{dx}) is identified with~$\Gamma^{-2}$. Moreover, the symmetry
condition for the 3$\times$3 matrix, equation (\ref{cond2}), is the
same as the gauge condition (\ref{symcond}) that we had to impose on
the spatial vielbeins in order to compute the matrix square root in
the solution for $D$.  

The effective metric $G=a^2g+2ab\,g\sqrt{g^{-1}f}+b^2f$ in terms of
the vierbeins simply becomes, 
\begin{align}
G_{\mu\nu}&=a^2 e_\mathrm{t}(g)^a_{~\mu}\,\eta_{ab}\, e_\mathrm{t}(g)^b_{~\nu}
+b^2 e_\mathrm{t}(f)^a_{~\mu}\,\eta_{ab}\, e_\mathrm{t}(f)^b_{~\nu}\nn\\
&~~~+ ab \Big[e_\mathrm{t}(g)^a_{~\mu}\,\eta_{ab}{\Lambda^b}_c \,e_\mathrm{t}(f)^c_{~\nu}+ e_\mathrm{t}(f)^a_{~\mu}\,\eta_{ab}{\Lambda^b}_c \,e_\mathrm{t}(g)^c_{~\nu}\Big]  
= e(G)^a_{~\mu}\,\eta_{ab}\, e(G)^b_{~\nu}\,,
\end{align} 
where the vierbein for $G$ is (modulo an overall Lorentz transformation),
\beqn\label{sumvier}
e(G)^a_{~\mu}=ae_\mathrm{t}(g)^a_{~\mu}+b{\Lambda^a}_b \,e_\mathrm{t}(f)^b_{~\mu}\,,
\eeqn
in agreement with~\cite{Noller:2014sta}.  

\subsection{Bound on the variables}

Since $v_a$ is a Lorentz velocity vector, we must have
$v_a\delta^{ab}v_b<1$ such that $\Gamma$ is finite. The symmetrization
conditions are therefore only solvable for the velocity vector if the
vierbein variables satisfy the bound, 
\beqn
\delta_{ab}{\big[(Ne^{-1}+L\varphi^{-1})^{-1}\big]^a}_i~(\nu^i-\lambda^i){\big[(Ne^{-1}+L\varphi^{-1})^{-1}\big]^b}_j~(\nu^j-\lambda^j)~<~1\,.
\eeqn
It is easy to see that the condition $v_a\delta^{ab}v_b<1$ translates
into $x>0$ which is the requirement for the existence of $\sqrt{x}$
and hence the square root-matrix~$\sqrt{g^{-1}f}$. This means that it
is possible to symmetrize the matrix $\hat{S}$ if and only
if $\sqrt{g^{-1}f}$ exists, as has already been pointed out
in \cite{Deffayet:2012zc} following a different approach.  

If the above bound is not satisfied, then bimetric theory does not
possess a formulation in terms of vielbeins. Whether the vierbein
formulation in this case is free of the Boulware-Deser ghost or not is
still an open question.

\section{Discussion}\label{discuss}

We have proven the absence of ghost at the classical level for a
recently proposed matter coupling in bimetric theory. The effective
metric that can consistently couple to matter is a combination of the
bimetric variables $g$ and $f$. Some relevant issues are discussed
below. 
 
Our ghost proof is important for the theoretical consistency of the
theory. As mentioned in the introduction, \cite{deRham:2014naa}
reported a perturbative analysis showing that the theory had a ghost
at the classical level. It was further argued that this ghost was
harmless for the low-energy theory because its mass was found to lie
above a certain cut-off scale. However, unlike healthy fields, ghosts
can create energy from the vacuum and thus they appear in interactions
with other particles, or in the process of vacuum decay, irrespective
of their mass or the available energy \cite{Woodard:2006nt,
  Sbisa:2014pzo}. A ghost mass above the cut-off scale of the theory
would imply that vacuum decay can spontaneously release energy above
the cut-off scale and render the effective description invalid.
Therefore the classical consistency of a theory requires the absence
of ghosts on all energy scales.

Quantum corrections may destabilize the specific structure of the
potential and/or the matter coupling and the Boulware-Deser ghost may
reappear \cite{deRham:2013qqa, deRham:2014naa}. This is not surprising
since already the formulation of a consistent quantum theory for
massive spin-1 fields requires the introduction of additional fields
and a Higgs mechanism. It is expected that a similar extension of
bimetric theory needs to be developed in order to achieve unitarity
also at the quantum level. The search for an analogue of the Higgs
mechanism for spin-2 fields is still on-going and, until it is found,
bimetric theory should be regarded as valid only at the classical
level.
 
Suppose that one of the parameters $a$ and $b$ in the effective metric
$G$ is fixed such that $b/a=m_f/m_g\equiv\alpha$. Then
$G=a^2\alpha(\alpha^{-1}g + g\sqrt{g^{-1}f}+\alpha f)$ is invariant
under the interchange of $\alpha^{-1}g$ and $\alpha f$. This
interchange symmetry becomes an invariance of the full theory if in
addition the $\beta_n$ parameters in the bimetric interaction
potential satisfy $\alpha^{4-n}\beta_n=\alpha^n\beta_{4-n}$. Models
with this specific symmetry have been discussed
in \cite{Hassan:2014vja}, where it was shown that they possess
solutions without a well-defined massive gravity limit. The presence
of the interchange symmetry at the classical level may have
implications for the quantum version of the theory because it is
expected that the symmetry is preserved at the quantum level provided
that it is compatible with the quantization procedure.
 
The fluctuations of the effective metric that respects the interchange
symmetry are massless around proportional backgrounds of bimetric
theory in vacuum \cite{Hassan:2012wr}. While the fluctuations of the
original metrics around maximally symmetric backgrounds are never mass
eigenstates, the parameters $a$ and $b$ that appear in the effective
metric (\ref{effm}) can be tuned to make its fluctuation massless or
massive. The coupling of a different effective metric with massless
fluctuations to matter has been studied before but turned out to
reintroduce the Boulware-Deser ghost \cite{Hassan:2012wr}. Coupling a
massless spin-2 field to matter is interesting because it avoids the
linear vDVZ discontinuity \cite{vanDam:1970vg, Zakharov:1970cc}. On
the other hand, the phenomenology of the theory could still differ
from general relativity because the effective metric does not possess
a standard kinetic term of Einstein-Hilbert form. Moreover, the
interaction with the massive spin-2 field may alter predictions for
observations. It is therefore interesting to study, for instance,
cosmological solutions and their perturbations in this new version of
bimetric theory including matter.
 
For most phenomenological applications, it is necessary to find
classical solutions to the equations of motion. The new ghost-free
matter coupling complicates the derivation of the equations for the
metrics $g$ and $f$: In order to compute the variation of the matter
coupling with respect to one of the metrics, it is necessary to know
the variation of the square root matrix $\sqrt{g^{-1}f}$. In
principle, this can be read off from the results
in~\cite{Guarato:2013gba} where the second variation of the bimetric
potential was computed. Another option is to switch variables from $g$
and $f$ to, for instance, $g$ and $G$, and compute the equations for
the new fields.\footnote{In~\cite{Hassan:2012wr}, the bimetric action
  has been rewritten in terms of $g$ and $S=\sqrt{g^{-1}f}$\,. From
  there, one can obtain the action in terms of $g$ and $G$ by
  replacing $S=\sqrt{g^{-1}G}-\mathbb{1}$\,.} In this case, the
variation of the matter coupling will be as simple as in general
relativity. The variation of the bimetric action becomes more
complicated but can still be computed in a straightforward way.  

\vspace{20pt}

\noindent
{\bf Acknowledgments:} We thank  C.~Deffayet, J.~Enander,
E.~M\"ortsell, B.~Sundborg and M.~von Strauss for helpful discussions. 

\appendix
\section{The sandwiched polar decomposition} \label{sandwich}
For matrices $A$ and $B$ and an orthogonal transformation $R$, such
that $R^\mT\JJ R=\JJ$, consider the matrix $ARB$.\footnote{The matrix
  $\JJ$ is defined in \eqref{idef} and we use this form of orthogonal
  rotations consistent with the conventions in this paper.} We
determine $R$ such that 
\beqn
ARB=(ARB)^\mT  
\eeqn
On inverting, using $R^{-1}=\JJ^{-1} R^\mT \JJ$ and manipulating the
outcome, this can be recast as,
\beqn
\left(A^\mT B^{-1} \JJ^{-1} R^\mT\right)^2=
(A^\mT B^{-1} \JJ^{-1})(A^\mT B^{-1} \JJ^{-1})^\mT\, \geq 0
\eeqn  
Hence the square-root of the right-hand-side exists. On taking the
square root and solving the right hand side for $R$ one gets the
desired result,
\beqn
R= \sqrt{(A^\mT B^{-1} \JJ^{-1})(A^\mT B^{-1} \JJ^{-1})^\mT}
\,\,(A^\mT B^{-1} \JJ^{-1})^{-1\mT}\,
\eeqn
\section{Alternative derivation of the effective lapse}\label{altder}
Another straightforward way to compute $N_{\mathrm{eff}}$ is directly
using (\ref{sumvier}), 
\begin{align}\label{app-eg1}
e(G)&=
\begin{pmatrix}N ~&~ 0\\
{e^a}_j\nu^j ~&~ {e^a}_i
\end{pmatrix}+\begin{pmatrix}\Gamma & \Gamma v_b\\
\Gamma v^a & \delta^{ac}\chi_{cb}
\end{pmatrix}\begin{pmatrix}L & 0\\
{\varphi^b}_j\lambda^j & {\varphi^b}_i
\end{pmatrix}\nonumber\\
&=\begin{pmatrix}N+L\Gamma+\Gamma v_b{\varphi^b}_j\lambda^j\,\, & \Gamma v_b{\varphi^b}_i\\
L\Gamma v^a+\delta^{ab}\chi_{bc}{\varphi^c}_j\lambda^j+{e^a}_j\nu^j\,\, & \delta^{ab}\chi_{bc}{\varphi^b}_i+{e^a}_i
\end{pmatrix}\,.
\end{align}
The matrix $\chi$ has been defined in~(\ref{chidef}) and again we raise indices on $v_a$ with $\delta^{ab}$.
We parametrize the vierbein $e(G)$ in terms of its block structure,
\begin{equation}
e(G)=\begin{pmatrix}e_{00} & e_{01}^{\tr}\\
e_{10} & e_{11}
\end{pmatrix},\label{app-eg2}
\end{equation}
where $e_{00}$ is a scalar, $e_{01}$ and $e_{10}$ are vectors and
$e_{11}$ is a spatial matrix. Then we can identify the lapse of the
corresponding metric, $G=e(G)^{\tr}\eta e(G)$, as, 
\begin{equation}
N_{\mathrm{eff}}^{2}=\left(e_{10}^{\tr}e_{11}-e_{00}e_{01}^{\tr}\right)\left(e_{11}^{\tr}e_{11}-e_{01}e_{01}^{\tr}\right)^{-1}\left(e_{11}^{\tr}e_{10}-e_{00}e_{01}\right)+e_{00}^{2}-e_{10}^{\tr}e_{10}\,.
\end{equation}
After invoking the matrix inversion lemma and performing some algebra, this simplifies
to,
\begin{equation}
N_{\mathrm{eff}}=\frac{e_{00}-e_{01}^{\tr}e_{11}^{-1}e_{10}}{\sqrt{1-\big|(e_{11}^{\tr})^{-1}e_{01}\big|^{2}}}\,,
\end{equation}
where we use the notation $\left|u\right|$ for the norm of the vector $u$ with respect to $\delta_{ab}$. Identifying
the components of (\ref{app-eg1}) with (\ref{app-eg2}) and substituting
$\nu^i=\lambda^i+{(Ne^{-1}+L\varphi^{-1})^i}_a\delta^{ab}v_b$,
we obtain, 
\begin{eqnarray}
e_{00} & = & N+L\Gamma+\Gamma v^{\tr}\varphi\lambda,\\
e_{01}^{\tr} & = & \Gamma v^{\tr}\varphi,\\
e_{10} & = & L(\JJ^{-1}\chi+ e\varphi^{-1})\JJ^{-1}v+N\JJ^{-1}v+(\JJ^{-1}\chi+e\varphi^{-1})\varphi\lambda,\\
e_{11} & = & (\JJ^{-1}\chi+e\varphi^{-1})\varphi\,,
\end{eqnarray}
where, as before, the matrix $\JJ^{-1}$ has components $\delta^{ab}$.
From this we can evaluate the numerator and denominator of $N_{\mathrm{eff}}$,
\begin{eqnarray}
e_{00}-e_{01}^{\tr}e_{11}^{-1}e_{10} & = & L\Gamma^{-1}+N\left(1-v^{\tr}(\JJ+\JJ e\varphi^{-1}\chi^{-1}\JJ )^{-1}v\right),\\
\sqrt{1-\left|(e_{11}^{\tr})^{-1}e_{01}\right|^{2}} & = & \sqrt{1-\left|((\JJ+\JJ e\varphi^{-1}\chi^{-1}\JJ)^{\tr})^{-1}v\right|^{2}}.
\end{eqnarray}
After assembling, we obtain, 
\begin{equation}
N_{\mathrm{eff}}=c_{1}N+c_{2}L,
\end{equation}
\begin{eqnarray}
c_{1} & = & \frac{1-v^{\tr}(\JJ+\JJ e\varphi^{-1}\chi^{-1}\JJ )^{-1}v}{\sqrt{1-\left|((\JJ+\JJ e\varphi^{-1}\chi^{-1}\JJ)^{\tr})^{-1}v\right|^{2}}},\\
c_{2} & = & \frac{\Gamma^{-1}}{\sqrt{1-\left|((\JJ+\JJ e\varphi^{-1}\chi^{-1}\JJ)^{\tr})^{-1}v\right|^{2}}}\,.
\end{eqnarray}
where it can be verified that $\left|((\JJ+\JJ e\varphi^{-1}\chi^{-1}\JJ)^{\tr})^{-1}v\right|^{2}<1\,$. 
The expression for $c_{1}$ can be further simplified to,
\begin{equation}
c_{1}=\frac{\Gamma^{-1}}{\sqrt{1-\left|((\JJ+\JJ\varphi e^{-1}\chi^{-1}\JJ)^{\tr})^{-1}v\right|^{2}}},
\end{equation}
which, when compared to $c_{2}$, reflects the symmetry of the equations with respect to $e(g)$ and~$e(f)$.  
Finally, we have verified that these expressions are in agreement with our result for $N_\mathrm{eff}$ in~(\ref{efflap}).


\end{document}